# Quantum magnetic resonance microscopy

David A. Simpson[1,2], Robert G. Ryan[3], Liam T. Hall[1], Evgeniy Panchenko[1], Simon C. Drew[4], Steven Petrou[2,4,5,6], Paul S. Donnelly[3], Paul Mulvaney[3], Lloyd C. L. Hollenberg[1,2,3,7]

[1]School of Physics, University of Melbourne, Parkville, 3010, Australia
[2]Centre for Neural Engineering, University of Melbourne, Parkville, 3010, Australia
[3]School of Chemistry and Bio21 Molecular Science and Biotechnology Institute, The University of Melbourne, VIC 3010, Australia
[4]Department of Medicine, Royal Melbourne Hospital, The University of Melbourne, Parkville, 3010, Australia
[5]Florey Neuroscience Institute, University of Melbourne, Parkville, 3010, Australia
[6]Centre for Integrated Brain Function, University of Melbourne, Parkville, Victoria, Australia
[7]Centre for Quantum Computation and Communication Technology, University of Melbourne, Parkville, 3010, Australia
*corresponding authors simd@unimelb.edu.au & lloydch@unimelb.edu.au

**ABSTRACT**

Magnetic resonance spectroscopy is universally regarded as one of the most important tools in chemical and bio-medical research. However, sensitivity limitations typically restrict imaging resolution to length scales greater than 10 μm. Here we bring quantum control to the detection of chemical systems to demonstrate high resolution electron spin imaging using the quantum properties of an array of nitrogen-vacancy (NV) centres in diamond. Our quantum magnetic resonance microscope selectively images electronic spin species by precisely tuning a magnetic field to bring the quantum probes into resonance with the external target spins. This provides diffraction limited spatial resolution of the target spin species over a field of view of ~50×50 μm$^2$. We demonstrate imaging and spectroscopy on aqueous $Cu^{2+}$ ions over microscopic volumes (0.025 μm$^3$), with detection sensitivity at resonance of $10^4$ spins/voxel, or ~100 zeptomol ($10^{-19}$ mol). The ability to image, perform spectroscopy and dynamically monitor spin-dependent redox reactions and transition metal biochemistry at these scales opens up a new realm of nanoscopic electron spin resonance and zepto-chemistry in the physical and life sciences.

Magnetic resonance spectroscopy techniques have revolutionised detection and imaging capabilities across the life and physical sciences. Electron spin resonance (ESR), nuclear magnetic resonance (NMR) and magnetic resonance imaging (MRI) are now essential tools in many areas of science and clinical research. State-of-the-art ambient NMR [1] and ESR-based systems [2] employing field gradients have demonstrated imaging resolution as low as 10 μm. However, exploring nanoscale biological and chemical processes with sub-micron requires a major technological shift.

Conventional ESR-based imaging approaches demonstrated detection from as few as $10^4$ spins with sub-micron spatial resolution, by reducing the size of the surface loop and scanning, however such measurements require cryogenic temperatures [3]. Other high resolution imaging techniques such as magnetic resonance force microscopy [4] and scanning tunnelling microscopy provide single-electron spin sensitivity with nanoscale spatial resolution [5], but they are also constrained to cryogenic temperatures and high vacuum environments which precludes their use in imaging functional bio-chemical reactions. For electron spin resonance applications, the regime of sub-micron room temperature spectroscopy and imaging has presented a major challenge. The development of such a technology would provide a fundamentally new view of electron spin dynamics at the nanoscale, including redox dynamics, organic radical formation and complicated transition metal biochemistry on a sub-cellular scale.

Here we report a new imaging technique based on quantum magnetic resonance spectroscopy, which can selectively image spectrally-resolved spin targets in aqueous solution with high spatial resolution under ambient conditions. Electron spin detection is achieved using an array of nitrogen vacancy (NV) spin probes in diamond together with wide-field optical microscopy and precise magnetic field alignment as shown in Fig. 1a. As a demonstration of the capabilities, we perform spectroscopic imaging of hexaaqua-$Cu^{2+}$ complexes, see Fig. 1b, and their associated redox dynamics over fields of view of ~50×50 $\mu m^2$. We demonstrate imaging resolution at the diffraction limit (~300 nm) with spin sensitivities in the zeptomol ($10^{-21}$) range.

**RESULTS**

Quantum magnetic resonance microscopy (QMRM) is performed using an array of solid state atomic sized quantum sensors in diamond. The NV defect sensors are engineered at control depths (~6-8 nm) from the diamond surface with a spatial separation on average of ~10 nm via nitrogen ion-implantation, see Methods. Due to the crystal symmetry of diamond, the quantum probes are formed along four possible <111> crystallographic orientations. The NV centre in diamond has a paramagnetic ground state triplet as shown in Fig. 1c, with the $|\pm 1\rangle$ states separated from the $|0\rangle$ state by the diamond zero-field splitting $D$ = 2.87 GHz. The degeneracy of the $|\pm 1\rangle$ states is lifted in the presence of a magnetic field and Zeeman split with a gyromagnetic ratio $\gamma_{NV}$ = 2.8 MHz/G. The quantum probes can be conveniently spin-polarised and their state "read out" at room temperature under optical excitation at 532 nm. The spin state readout arises from the difference in fluorescence intensity between the $|0\rangle$ and $|\pm 1\rangle$ states, allowing optically detected magnetic resonance (ODMR) of the ground state magnetic sublevels [6]. In the case of NV ensembles, the four orientations result in eight possible transition states depending on the applied magnetic field. An external field can be applied along a particular NV symmetry axis and a single microwave π pulse used to isolate the subset of NV spin probes which are aligned with the applied magnetic field. The spin lattice relaxation time ($T_1$) of the quantum probes can be determined by optically polarizing the spins into the $m_s$= 0 ground state, then allowing them to evolve for a time τ, before sampling the spin population with an additional optical pulse. Interactions between the NV probe and nearby electronic, nuclear and surface spins species cause the NV net magnetization to relax from the $m_s$ = 0 state to a mixture of the three ground triplet states. The 1/e decay time is referred to as the $T_1$ time of the probe.

The NV centre [7] in diamond is a promising system for nanoscale electronic spin detection ideally suited for room temperature biological environments [8-10]. While detection of small numbers of electronic spins, based on decoherence [11], or quantum control protocols [12] has been carried out using single NV centres [13-17], and in wide-field array systems [18, 19], spectroscopic microscopy that can target specific electronic spin species is lacking. Spectroscopic detection to date has relied on complex quantum control techniques [20, 21], which creates major challenges when scaling up to large fields of view. We instead employ an optical spectroscopic detection method [22] based on precise control of a static external magnetic field, $B_0$, which is used to tune the NV transition frequency into resonance with environmental spins at selected effective g-factors. At a given resonance condition the NV and environmental spins (which may comprise intrinsic and target spins) exchange energy resulting in an increase in the measured NV relaxation rate as illustrated in Fig. 1d.

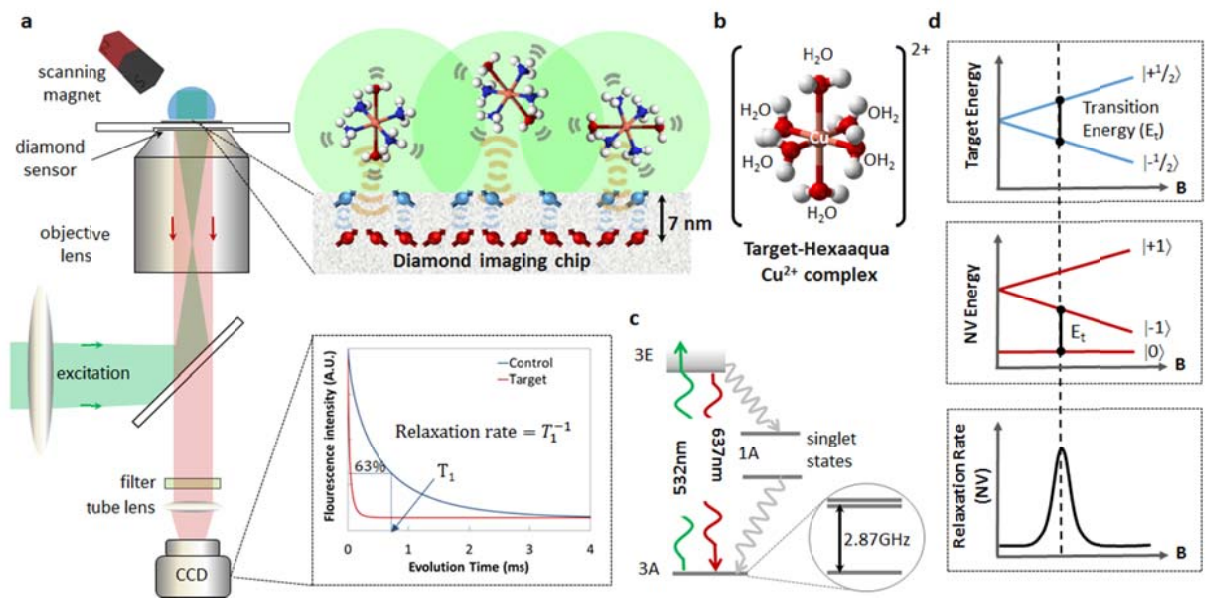

Figure 1. Quantum magnetic resonance microscopy **a**, Schematic diagram of the quantum magnetic resonance microscope. Optical excitation is performed with 532 nm light and the resulting NV fluorescence is filtered (650-750nm) and imaged onto an sCMOS camera. Microwave control is achieved using a gold microwave resonator evaporated onto a glass coverslip. Expanded view on the right shows the diamond imaging chip with the 2D layer of NV centres ~10 nm below the surface. The spin lattice relaxation time, $T_1$, from the imaging array is determined from a sequence of fluorescence images, see Fig. 2 for more details. **b**, A schematic representation of the spin target structure of hexaaqua-$Cu^{2+}$, $[Cu(OH_2)_6]^{2+}$ with the $Cu^{2+}$ ion in a distorted octahedral environment. **c**, The simplified energy level diagram for the NV centre highlighting the paramagnetic ground state triplet. At zero magnetic field the degenerate $|-1\rangle$ and $|+1\rangle$ states are separated from the $|0\rangle$ state by the diamond crystal field splitting of $D$ = 2.87 GHz. **d**, Electron spin spectroscopy technique. The two simplified energy diagrams represent the Zeeman splitting as a function of B field for a target spin ($Cu^{2+}$) and the NV. When the transition energy from the target spin ($|-^1/_2\rangle \rightarrow |+^1/_2\rangle$) is matched to the transition energy from the NV centre ($|0\rangle \rightarrow |-1\rangle$), the target and NV spins exchange energy efficiently, resulting in a reduction in the spin relaxation time ($T_1$) (increase in the rate) of the NV centre. The cross relaxation occurs within the sensing volume depicted by the green shaded regions, which in practice extend to approx. 10 nm.

The relaxation rate, $\Gamma_1^{(E)}(B_0) = 1/T_1$, of a quantum probe in the presence of an arbitrary magnetic environment (E), comprising intrinsic (I) and/or target (T) spins, is given by the spectral density of the environment $S^{(E)}(\omega_E, B_0)$ convolved with the filter function of the probe, in this case the NV spin:

$$\Gamma_1^{(E)}(B_0) = \int S^{(E)}(\omega_E, B_0) G(\Gamma_2, \omega_E, B_0) \, d\omega_E , \qquad (1)$$

where $G(\Gamma_2, \omega_E, B_0) = \frac{b^2 \Gamma_2}{2(\Gamma_2^2 + (\omega_{NV} - \omega_E)^2)}$ is the NV filter function given by a Lorentzian dependent on the resonance frequency of environmental spins, $\omega_E$, the coupling strength, $b$, the external magnetic field $B_0$ ($\omega_{NV} = \gamma_{NV} B_0$) and NV transverse relaxation rate $\Gamma_2$, see SI. To probe specific components of the environment's spectral density, $S^{(E)}(\omega_E, B_0)$, The NV filter function can be tuned via the Zeeman interaction and an applied magnetic field, $B_0$.

To characterise the inherent (I) magnetic environment of the diamond sensing chip a calibration experiment is carried out to obtain the relaxation rate spectrum, $\Gamma_1^{(I)}(B_0)$ in the absence of a target system. The calibration spectrum reflects the interactions of electronic and nuclear spins both on the surface and within the bulk diamond. The target (T) is then introduced and the NV relaxation rate spectrum corresponding to the combined system, $\Gamma_1^{(I+T)}(B_0)$, mapped. By subtracting the calibration measurement we obtain the target relaxation spectrum $\Gamma_1^{(T)}(B_0) \approx \Gamma_1^{(I+T)}(B_0) - \Gamma_1^{(I)}(B_0)$, see SI for details. Given the relatively narrow probe filter function, the target relaxation spectrum is essentially unchanged following deconvolution of (1) and is therefore identical to the spectral density $S^{(T)}(\omega_E; B_0)$, up to a normalization factor.

## 1. Quantum magnetic resonance imaging of hexaaqua-$Cu^{2+}$ ions

Our target system, the hexaaqua-$Cu^{2+}$ complex, is a bio-essential trace nutrient present in many enzymes. The incorporation of copper into metalloproteins is tightly controlled in biological systems due to the redox activity of free copper ions which results in the production of reactive free radical species which cause cellular damage [23-25]. Non-invasive imaging of the copper distribution is therefore a key aspiration that will ultimately provide direct information on cellular metabolism. To date the majority of techniques looking to image $Cu^{2+}$ have focused on the development of species specific target molecules [26] which bind $Cu^{2+}$ complexes resulting in an increased fluorescence signal or changes in the fluorescence lifetime of the target molecule [27]. The quantum magnetic resonance approach described here is non-invasive and does not impact on the function or availability of the $Cu^{2+}$ complexes. This is an important distinction for biological applications which aim to understand the role $Cu^{2+}$ plays in cell signaling.

We begin with the demonstration of selective detection and spatial imaging of $Cu^{2+}$ ions. Figure 2a shows the transition frequencies of the NV probe, free electron spins ($\langle g \rangle = 2$) and the target hexaaqua $Cu^{2+}$ ion ($\langle g_{eff} \rangle = 2.199$) as a function of magnetic field. The electronic configuration of the target spin complex [Cu(OH$_2$)$_6$]$^{2+}$, is d$^9$ with octahedral geometry, giving rise to a single unpaired electron, and an overall electronic spin of ½. Degeneracy resulting from Jahn-Teller distortion, see Fig. 1b, gives rise to an axially-symmetric Zeeman interaction: $g_\perp$=2.099 and $g_{//}$=2.400 [28]. At room temperature a weighted average of these two g factors is observed due to motional averaging; when integrated over all possible orientations this results in an in an isotropic g-factor of $\langle g \rangle = 1/3(g_{||} + 2g_\perp)$ = 2.199. From Fig. 2a the resonance condition for Cu$^{2+}$ ions ↔ NV occurs around 487 G. At the resonance point the Cu$^{2+}$ and NV probe spins can exchange energy efficiently resulting in a significant reduction in the NV T$_1$ time (increase in the relaxation rate), forming the fundamental QMRM image contrast.

Figure 2b presents two T$_1$ relaxation curves (measurement sequence shown in the inset) off (250 ± 1 G) and near resonance (460 ± 2 G) with Cu$^{2+}$ spins. The T$_1$ time from the NV probes reduced from 730 ± 60 µs to 96 ± 1 µs as the Cu$^{2+}$ resonance condition was approached. This reduction included a contribution from surface spins, the analysis and subtraction of which is described quantitatively in the next section on spectroscopy. To image the Cu$^{2+}$ distribution in solution an image mask was fabricated on top of the diamond imaging chip with Poly(methyl methacrylate) (PMMA) as described in Fig. 2c, see Methods. A set of single $\tau_{sp}$ images were obtained at magnetic fields either side of the Cu$^{2+}$ ↔ NV resonance position. The $\tau_{sp}$ probe time was set to the T$_1$ time obtained from the field of view (FOV) at each respective magnetic field. This probe time maximized image contrast for small signal changes. The single $\tau_{sp}$ images were averaged over $N_c$ = 6×10$^6$ cycles and are shown in Fig. 2d.

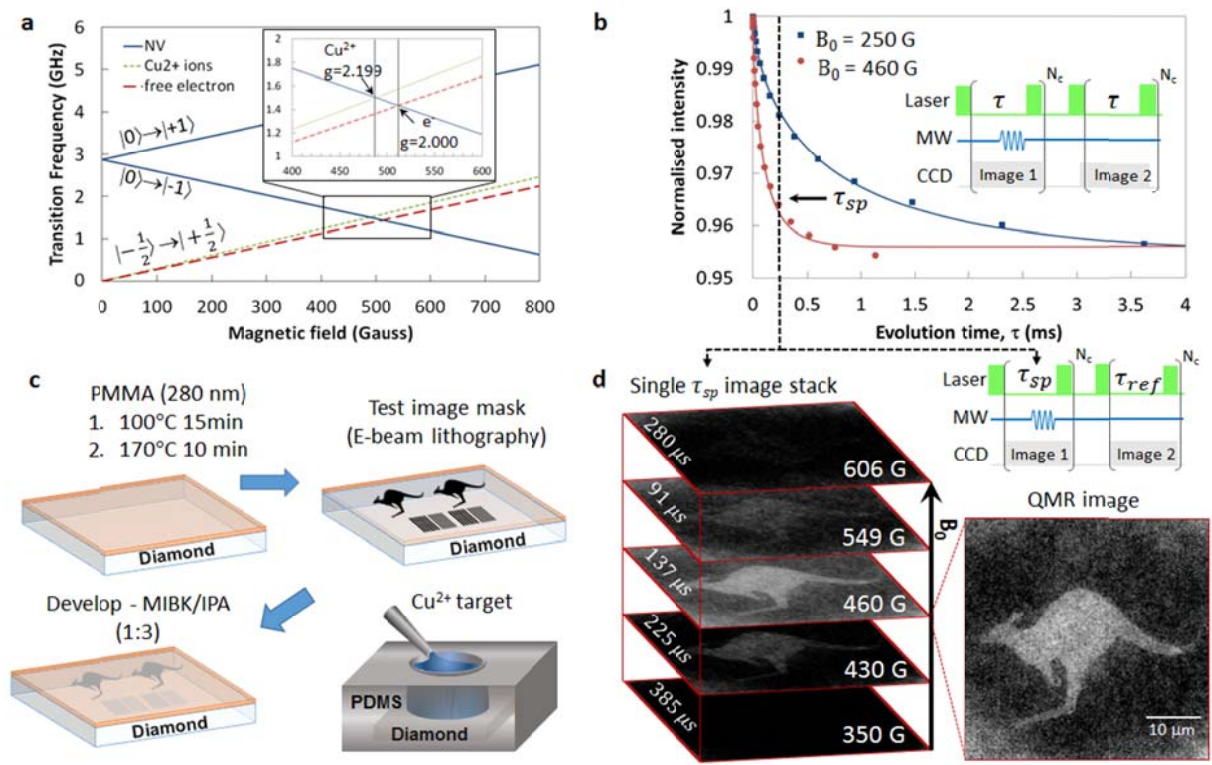

Figure 2. Quantum magnetic resonance imaging of aqueous Cu$^{2+}$ – a, Transition frequencies of the hexaaqua-Cu$^{2+}$ ions ($|-^1/_2\rangle \rightarrow |+^1/_2\rangle$) and NV spin probes ($|0\rangle \rightarrow |-1\rangle$ and $|0\rangle \rightarrow |+1\rangle$) as a function of the applied B-field.

**Inset:** Zoomed plot of the resonant positions of NV and Cu$^{2+}$ (487 G) as well as NV and free electron spins (512 G). **b**, T$_1$ relaxation curves integrated over the imaging area 10×10 μm on and off resonance with Cu$^{2+}$ spins. On-resonance data was taken at 460 ± 2 G. Off-resonance data is taken at 250 ± 2 G. Inset shows the T$_1$ measurement sequence. Measurements were averaged over $N_c$ = 10,000 cycles at each 'τ' point, with a typical acquisition time of 5 minutes per T$_1$ curve. **c**, E-beam fabrication of an image mask using Poly(methyl methacrylate) (PMMA). **d**, Single $\tau_{sp}$ image stack at various magnetic field strengths through the target-probe resonant point. The measurement sequence is shown on the far right. The probe time, $\tau_{sp}$, at each magnetic field is shown on the left hand side of the image. Images were averaged over $N_c$ = 6 × 10$^6$ cycles. The total acquisition time per image was 40 minutes. The quantum magnetic resonance (QMR) image on the right shows clearly identifies the spatial distribution of Cu$^{2+}$ ions near the Cu$^{2+}$ ↔ NV resonance at 460 G.

The quantum magnetic resonance (QMR) images show significant image contrast between areas exposed to the Cu$^{2+}$ solution defined by the image mask at magnetic fields near the Cu$^{2+}$ ↔ NV resonance condition. This form of imaging is sufficient to spatially image the Cu$^{2+}$ distribution in solution. In the following section we perform quantitative spectroscopy on these target spins.

## 2. Quantitative electronic spin spectroscopy

To demonstrate quantitative spectroscopy, a thorough understanding of the intrinsic magnetic environment consisting of the surface and intrinsic spins is required before introducing external target spins into the environment. Calibration measurements were conducted with a solution of nitric acid (4 μL, 1 mM), which was used for dissolution of the Cu$^{2+}$ analyte. The calibration spectrum, $\Gamma_1^{(I)}(B_0)$, is shown in Fig. 3a (green diamonds). The strength of the external B-field was determined directly from the Zeeman splitting of the NV $|\pm 1\rangle$ energy states. The magnetic field gradient over the FOV was < 0.4%, see SI. The calibration spectrum was described well by a single Lorentzian fit. The peak at 511.3 ± 0.5 G corresponds to interactions between the NV probes and surface electronic spins with an average $\langle g \rangle$ = 2.01 ± 0.02. The half-width of the calibration spectrum ω = 60.4 ± 3 × 10$^6$ rad s$^{-1}$ indicates that the dominant contribution to the linewidth is a spin relaxation process on the time scale of tens of ns, consistent with surface spin phonon relaxation being the dominant contribution to the intrinsic spectrum [29]. A surface spin density of 2.4 spins/nm$^2$ can be estimated from the intrinsic spectrum, see SI. Our method thus provides rapid and quantitative spectroscopy of the surface spin noise spectrum and can be used in combination with a variety of surface chemical passivation modes in order to understand and mitigate these effects.

With the intrinsic calibration spectrum $\Gamma_1^{(I)}(B_0)$ obtained, quantitative spectroscopy can be performed on the spin target by acquiring the combined T$_1$ relaxation rate spectrum, $\Gamma_1^{(I+Cu)}(B_0)$. Figure 3a (blue circles) presents the combined spectrum from a 4 μl (100 mM) droplet of Cu$^{2+}$ solution. The Cu$^{2+}$ spectrum, $\Gamma_1^{(Cu)}(B_0)$ is then obtained by simply subtracting the calibration spectrum as shown in Fig. 3b. The Cu$^{2+}$ spectrum is de-convolved in the SI using the measured NV filter function, however since the width of the filter function ~ 4 MHz ($\Delta\omega$ = 25.1 × 10$^6$ rad s$^{-1}$) is significantly less than the half linewidth of the Cu$^{2+}$ spectrum ($\Delta\omega$ = 1.8 × 10$^9$ rad s$^{-1}$) the relaxation rate spectrum itself represents the spectral density of the hexaaqua-$Cu^{2+}$ environment. For comparison we have plotted the microwave absorption spectrum obtained from conventional continuous wave (CW)-ESR spectrometer. The plots in Fig. 3b were independently normalised to the area under the spectrum and show excellent agreement between the QMRM spectrum from the FOV (red circles) and the conventional CW-ESR spectrum. We emphasise here that the hexaaqua-$Cu^{2+}$ spectral density is obtained by simply bringing the target spin into resonance with no active driving of the target spin.

The $Cu^{2+}$ spectrum can be derived theoretically and is given by the following expression, see SI:

$$\Gamma_1^{(Cu)}(B_0) = \frac{b_{Cu}^2}{2} \frac{\Gamma_2^{(I+Cu)}+R_{Cu}}{(\Gamma_2^{(I+Cu)}+R_{Cu})^2+(2\pi D-(\gamma_{NV}+\gamma_{Cu})B_0)^2} \,, \qquad (2)$$

where the $\gamma_{Cu}$ is the gyromagnetic ratio of the $Cu^{2+}$ spins, $b_{Cu}$ characterises the strength of the overall probe-target coupling (see SI), and $R_{Cu}$ is the total width of the spectral density arising from various processes intrinsic to the $Cu^{2+}$ solution, e.g. intrinsic relaxation ($R_{Cu}^{Relax}$), dipole-dipole interactions ($R_{Cu}^{dip}$), spatial ($R_{Cu}^{Spatial}$) and rotation motion ($R_{Cu}^{Rot}$). Our analysis shows that the dominant contribution to the linewidth arises from the intrinsic fluctuation rate, i.e. $R_{Cu} \approx R_{Cu}^{Relax}$ which is concentration independent and of order GHz (see SI for further details) [28].

By fitting Eq. (2) to the $Cu^{2+}$ spectrum, the depth of the NV probes, $h_{probe}$, and the $Cu^{2+}$ intrinsic fluctuation rate, $R_{Cu}^{Relax}$ can be obtained. The fit to the theoretical model, shown in Fig. 3b, is in excellent agreement with the data and yields an estimated probe depth of $h_{probe}$ = 6.7 ± 0.04 nm and $R_{Cu}^{Relax}$ = (1.8 ± 0.05) × $10^9$ rad $s^{-1}$. The NV depth is consistent with molecular dynamics simulations [30]. The $Cu^{2+}$ intrinsic fluctuation rate, $R_{Cu}^{Relax}$ is in excellent agreement with the value obtained from bulk ESR of $R_{Cu}^{Relax} = (1.800 \pm 0.005) \times 10^9$ rad $s^{-1}$, validating our experiment and theoretical approach. The peak position of the spectrum ($B_{res}$ = 486 ± 1 G) is directly related to the g-factor of $Cu^{2+}$ spins by $\langle g_{Cu} \rangle = \frac{h\,D}{\mu_B\,B_{res}} - g_{NV}$, where is $h$ Planck's constant, $D$ is the zero field splitting of diamond, $\mu_B$ is the Bohr magneton, $g_{NV}$ is the effective g-factor of the NV spin 2.0028 [7] and $B_{res}$ is the magnetic field value corresponding to the peak of the spectrum. From the theoretical fit the measured value of $\langle g_{Cu} \rangle = 2.21 \pm 0.02$ is in excellent agreement with the literature [31].

The spectral resolution of our technique is ultimately limited by the point spread function (PSF) of the NV filter function. The width of the PSF is defined by hyperfine and free induction decay mechanisms ($\Gamma_2$). For the NV array used in this work the PSF is ~ 4 MHz or 0.14 mT. Improvements such as isotopically pure $^{12}C$ diamond would improve this resolution down to a few hundred kHz [9]. In the following section we quantitatively investigate the spectroscopic electronic spin imaging via QMRM.

## 3. High resolution electron spin imaging and spectroscopy

The $Cu^{2+}$ spectrum shown in Fig. 3b was obtained by integrating the signal over the entire FOV at each applied magnetic field. To investigate the spectroscopic spatial resolution of the quantum magnetic resonance microscope we performed a separate experiment, and interrogate a subset of pixels. The goal was to determine the $T_1$ relaxation rates of the NV probes from a target sensing volume defined by the imaging pixel size and probe depth. The relaxation maps, $T_1(B_0: x,y)$, for the calibration and target spin solutions were subtracted to give a set of $Cu^{2+}$-induced relaxation images, $\Gamma_1^{(Cu)}(B_0: x,y)$, as a function of magnetic field strength as shown in Fig. 3c. The $Cu^{2+}$ spectrum was then extracted from a region of interest (ROI) of $1.6 \times 1.6$ μm$^2$. Figure 3b (green circles) shows the ROI $Cu^{2+}$ spectrum, which is in excellent agreement with the previous spectrum obtained over the entire FOV. To investigate the uniformity of the measured signal across the imaging area we plot the measured g-factor from each ROI pixel as a histogram in Fig. 3d. The distribution is uniform over the imaging area with a mean of $\langle g_{Cu} \rangle = 2.19 \pm 0.03$, consistent with the results for the FOV. A similar procedure was conducted for the calibration spectrum which yields a narrow distribution centered at $\langle g_{surface} \rangle = 2.003 \pm 0.007$ consistent with the g-factor for free electrons, see Fig. 3d.

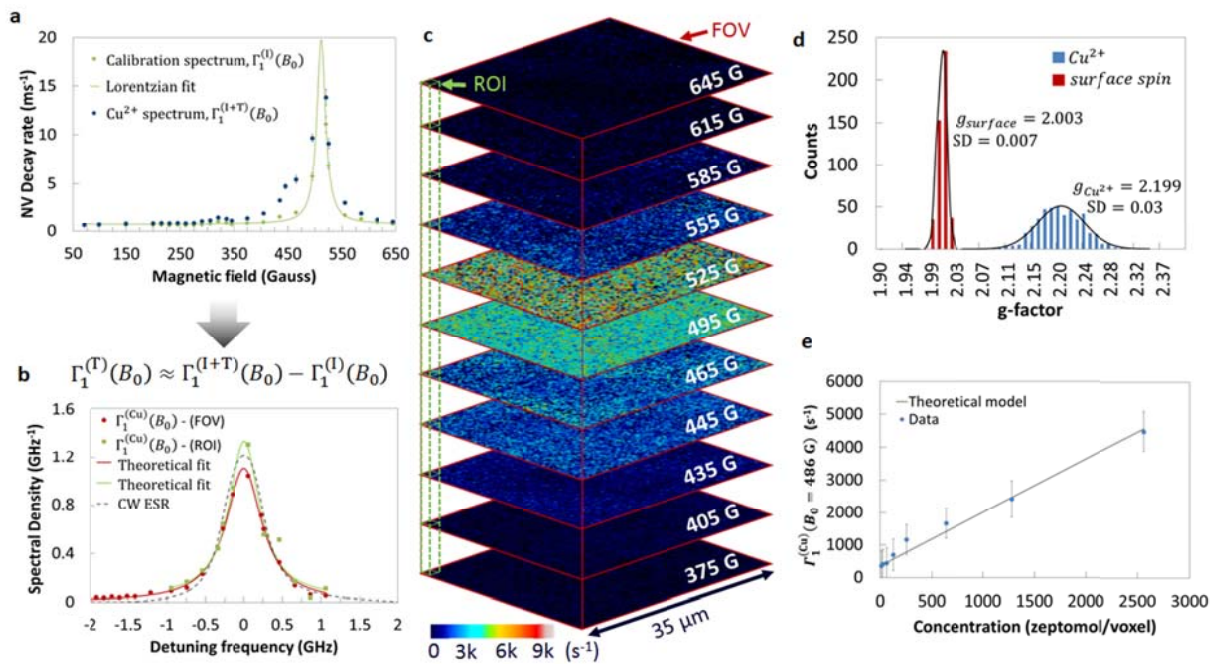

Figure 3: High resolution electronic spin imaging and spectroscopy. **a,** Measured NV decay rate as a function of $B_0$ for 1 mM of nitric acid and 100 mM of $Cu^{2+}$ ions. The calibration spectrum (green squares) is described well with a Lorentzian fit with a peak at $511.3 \pm 5$ G and half linewidth of $\omega = 60.4 \pm 3 \times 10^6$ rad s$^{-1}$. **b,** Relaxation rate spectra, $\Gamma_1^{(Cu)}(B_0)$, from the FOV and ROI defined in **c,** from 100 mM hexaaqua-$Cu^{2+}$ at room temperature. The absorption peak at $487 \pm 4$ G corresponds to a g-factor of $2.20 \pm 0.04$ which is consistent with the effective g-factor for aqueous $Cu^{2+}$ ions from conventional CW ESR see dashed gray line. **c,** $Cu^{2+}$-induced relaxation rate images, $\Gamma_1^{(Cu)}(B_0: x,y)$, as a function of applied magnetic field. The pixel size in **c,** is $400 \times 400$ nm. To obtain the spectral density/voxel the relaxation rates were binned 4×4 to generate the ROI spectrum from a pixel voxel size of $1.6 \times 1.6 \times 0.01$ μm. A typical ROI $Cu^{2+}$ spectrum from a single voxel is shown in solid green in **b**. **d.** Histogram of the measured g-factor from $Cu^{2+}$ and surface spins obtained from the respective relaxation

spectra/voxel. The distributions reveal an average g-factor of 2.003 for the surface spins and 2.199 for the $Cu^{2+}$ ions as expected. **e,** Measured relaxation rate, $\Gamma_1^{(Cu)}$, on resonance ($B_0$ = 486 G) versus $Cu^{2+}$ concentration. The minimum detectable concentration is 5 mM which in a sensing volume defined by 0.025 $\mu m^3$ or (25 aL) corresponds to the detection of ~ 75,000 spins/voxel or ~125 zeptomol/voxel.

At this point, we quantify the number of detected spins per sensing volume. The ROI sensing voxel is defined by 1.6×1.6×0.01 $\mu m^3$ (0.025 $\mu m^3$) or 25 aL, see SI. For 100 mM $Cu^{2+}$ this equates to the detection of 1.5×10$^5$ $Cu^{2+}$ spins or ~2 attomol per voxel. This is by no means the detection limit. To determine the sensitivity of the system the concentration of $Cu^{2+}$ ions in solution was varied whilst on resonance with the NV probes, as shown in Fig. 3e. Using the NV probe height and intrinsic $Cu^{2+}$ fluctuation rate, $R_{Cu}^{Relax}$ obtained from the $Cu^{2+}$ spectrum (FOV) we can fit the concentration data using the following expression, see SI:

$$[Cu^{2+}] \,[mol/L] = \frac{4.35\times10^{10}\cdot\Gamma_1^{(Cu)}(B_{res})\cdot h_{probe}^3\left(\Gamma_2^{(I+Cu)}+R_{Cu}^{Relax}\right)^2}{\left(\Gamma_2^{(I+Cu)}+R_{Cu}^{Relax}\right)}, \quad (3)$$

where, $\Gamma_1^{(Cu)}(B_{res})$ is the decay rate from the $Cu^{2+}$ spins measured on resonance, $h_{probe}$ = 6.7 nm, $\Gamma_2^{(I+Cu)}$ = 4 MHz and $R_{Cu}^{Relax} = 1.8\times10^9$ rad s$^{-1}$. From the data the minimum detectable limit was 5 mM which for a single voxel corresponds to the detection of ~75,000 spins or ~125 zeptomol. In terms of spin sensitivity this represents a 10$^4$ improvement over current ambient ESR imaging systems.

To quantify the ultimate spatial imaging resolution we move to a separate region of the imaging chip with an image mask comprised of a series of gratings of width ~500 nm and pitch of 1 μm as shown in Fig. 4a. The QMR image at 460 G shown in Fig. 4b reveals the $Cu^{2+}$ spatial distribution. A line cut through a series of grating lines is shown in Fig. 4c, demonstrating an imaging resolution of 290 ± 30 nm which is in excellent agreement with the optical resolution of the microscope 1.22λ/(2NA) ≈ 305 nm.

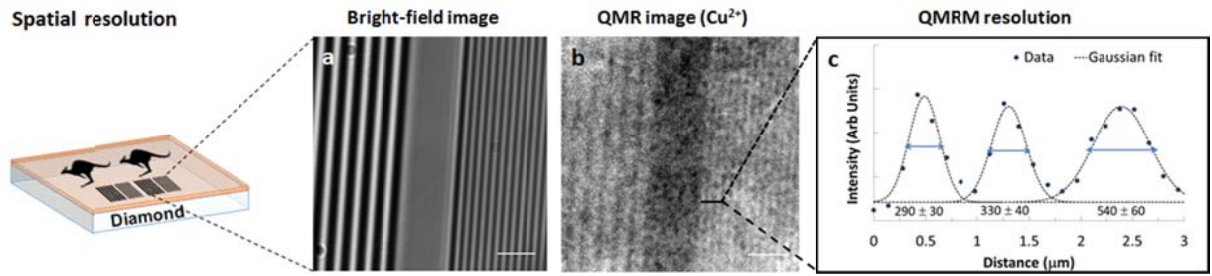

Figure 4: Spatial resolution of quantum magnetic resonance microscopy. **a,** Bright field image of the image mask used to determine the imaging resolution. **b,** Single $\tau_{sp}$ resonance image of $Cu^{2+}$ ions at 460 G ($\tau_{sp}$ = 124 μs, $n$ = 12×10$^6$). The light regions indicate areas of $Cu^{2+}$. **c,** Line cut through the a section of the quantum magnetic resonance image. The fit to the line scan demonstrates a diffraction limited resolution of 290 ± 30 nm.

These results show that spatial imaging of $Cu^{2+}$ ions with diffraction limited resolution and zeptomol sensitivity can be achieved under ambient conditions.

## 4. Monitoring redox reactions kinetics

Finally, we demonstrate dynamic spin detection by monitoring the redox state of $Cu^{2+}$ ions in the presence of a reducing agent, ascorbic acid ($AH_2$) as described in Fig 5a. To capture dynamic redox changes we implement a two-point $T_1$ measurement scheme with an external magnetic field set to $Cu^{2+} \leftrightarrow NV$ resonant point $B_0 = 486$ G, see Fig. 5b. The reaction kinetics for $Cu^{2+}$ in the presence of ascorbic acid can be described by an anaerobic chain-beginning reaction $2Cu^{2+} + AH_2 \leftrightarrow 2Cu^+ + A + 2H^+$. The low value of the equilibrium constant, $K_e = (5 \pm 2) \times 10^{-9} M^2$ [32], suggests that after the initial reduction to $Cu^+$ the $Cu^+$ re-oxidizes back to $Cu^{2+}$ which dominates at long times, consistent with the expected disproportionation of $Cu^{2+}$ ions.

Figure 5c (red squares) shows a control measurement performed by diluting the $Cu^{2+}$ 100 mM solution by 1:1 with MilliQ water. The integrated fluorescence intensity from the FOV was recorded at 3 sec intervals over 4000 seconds. The $t = 0$ point represents the time MilliQ water was added. The fluorescence intensity change is consistent with the change expected for $Cu^{2+}$ concentration reducing from 100 to 50 mM. The measured fluorescence intensity is proportional to the $Cu^{2+}$-induced decay rate and is given by Eq. (4), see SI for details:

$$\Gamma_1^{(Cu)}(B_{res}) = \frac{\Delta I}{c\, \tau_{sp} \Delta I_0}, \qquad (4)$$

where $\tau_{sp}$ is the measurement time point in the two-point $T_1$ measurement sequence, $\frac{\Delta I}{\Delta I_0}$ is the percentage change in fluorescence intensity, $c = 0.04$ is the spin relaxation contrast governed by the NV ensemble.

Equations (3) and (4) can be used to translate the change in fluorescence intensity into the $Cu^{2+}$ concentration. This allows the $Cu^{2+}$ concentration per voxel to be determined over time as shown in Fig. 5c. To investigate dynamic redox reactions we introduce ascorbic acid ($AH_2$) at a ratio of 1:1 with $Cu^{2+}$ ions. Figure 5c (blue squares) shows the two point $T_1$ measurement from the FOV over the same time period as the control.

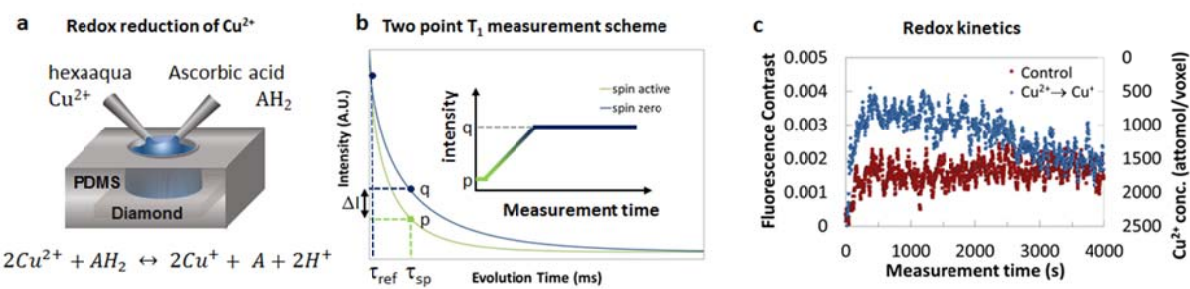

Figure 5: Monitoring redox kinetics. **a,** Schematic showing the redox reaction of $Cu^{2+}$ ions (spin ½) to cuprous ions (spin zero) via ascorbic acid at a ratio of 1:1. **b,** Dynamic quantum magnetic resonance microscopy using a two point $T_1$ measurement scheme to detect redox reactions of spin active molecules. The two time points are required to normalize the fluorescence signal over the entire FOV (50x50µm). **c,** Two point $T_1$ measurement showing the control experiment (red squares) consisting of a dilution of the $Cu^{2+}$ from 100 mM to 50 mM with water and the reduction of $Cu^{2+}$ using ascorbic acid (blue circles). The reduced $Cu^{2+}$ is found to disproportionate/re-oxidise over a period of 1 hour. The fluorescence intensity was determined from the FOV at measurement intervals of 3 sec.

The ascorbic acid was found to reduce ~ 60% of the $Cu^{2+}$ ions in the FOV within 500s. The reduced spin-zero cuprous ions are not stable in solution and slowly convert back into $Cu^{2+}$ via disproportionation and/or aerobic oxidation after 4000s. The final intensity returns to that of the diluted control concentration. This demonstration shows that the redox kinetics from sensing volumes of ~ 25 fL can be monitored over time and with a sensitivity of order attomol.

**DISCUSSION**

In summary we have demonstrated a new quantum magnetic resonance microscope using an array of quantum probes in diamond. The microscope operates at room temperature and in ambient conditions and allows simple non-invasive spectroscopy of unpaired electron spin systems in aqueous solution and under biologically compatible conditions. In demonstrating the system we have focused on the spectroscopy and imaging of the hexaaqua-$Cu^{2+}$ complex in aqueous solution. We have demonstrated species specific spatial imaging of the spin target with diffraction limited resolution at 300 nm and ultimate spin sensitivity approaching the zeptomol level. We have shown how quantitative spectroscopic imaging can be performed on external $Cu^{2+}$ spins in sensing volumes down to 25 aL with micron spatial resolution.  The theoretical framework describing the interaction of the target spin system and the NV spin probe is in excellent agreement with the experimental data. The application of quantum control to the detection and imaging of electronic spin systems represents a significant step forward. The work reported here demonstrates that quantum sensing systems can accommodate the fluctuating Brownian environment encountered in "real" chemical systems and the inherent fluctuations in the spin environment of ions undergoing ligand rearrangement. With high spin sensitivity and axial resolution, QMRM is ideal for probing fundamental nanoscale bio-chemistry such as binding events on cell membranes [33] and the intra-cellular transition metal concentration in the periplasm of prokaryotic cells [34].

**Methods**

*Materials*

The diamond imaging sensor used in this work is engineered from electronic grade Type IIa diamond (Element 6). The diamonds were thinned, cut and re-polished to a $2 \times 2 \times 0.1$ mm$^3$ crystal (DDK, USA). NV defects were engineered via ion implantation of $^{15}$N atoms at an energy of 4 keV and dose of $1 \times 10^{13}$ ions/cm$^2$. Molecular dynamic simulations indicate a NV depth range between 5-10 nm [30] which is consistent with the depth measurement from our analysis of 6.7 nm. The implanted sample was annealed at 1000 °C for three hours and acid treated to remove any unwanted surface contamination. The density of NV centres post annealing was $1 \times 10^{11}$ NV/cm$^2$. The linewidth of the ODMR was typically 4 MHz. The hyperfine spectrum from the $^{15}$N implant could not be resolved due to the inhomogeneous broadening from the dipole coupling of $^{13}$C spins present in the material at a concentration of 1.1%. The typical $T_1$ time of the array was 1.8 ms off resonance with a de-phasing time $T_2$ of ~2 μs.

*Optical Imaging*

The wide-field imaging was performed on a modified Nikon inverted microscope (Ti-U). Optical excitation from a 532 nm Verdi laser was focused (f = 300 mm) onto an acousto-optic modulator (Crystal Technologies Model 3520–220) and then expanded and collimated (Thorlabs beam expander GBE05-A) to a beam diameter of 10 mm. The collimated beam was focused using a wide-field lens (f = 300 mm) to the back aperture of the Nikon x60 (1.4 NA) oil immersion objective via a Semrock dichroic mirror (Di02-R561-25 × 36). The NV fluorescence was filtered using two bandpass filters before being imaged using a tube lens (f = 300 mm) onto a sCMOS camera (Neo, Andor). Microwave excitation to drive the NV spin probes was applied via an omega gold resonator (diameter=0.8mm) lithographically patterned onto a glass coverslip directly under the diamond imaging chip. The microwave signal from an Agilent microwave generator (N5182A) was switched using a Miniciruits RF switch (ZASWA-2-50DR+). The microwaves were amplified (Amplifier Research 20S1G4) before being sent to the microwave resonator. A Spincore Pulseblaster (ESR-PRO 500 MHz) was used to control the timing sequences of the excitation laser, microwaves and sCMOS camera and the images where obtained and analysed using custom LabVIEW code. The excitation power density used for imaging was 30 W/mm$^2$ and all images were taken in an ambient environment at room temperature.

*Image mask*

The fabrication of the image mask was done by cleaning the diamond imaging chip in hot acetone (65 °C) followed by rinsing in isopropanol alcohol (IPA) and deionised water. The sample was then spin-coated with 280 nm thick PMMA A4 950k resist. After the spin-coating the sample was baked on the hot plate for 15 min at 100°C followed by additional 10 min at 170°C. This enables a better solvent evaporation and prevents the resist from outgassing in vacuum. A 30 nm thick conduction layer of Cr was deposited on the resist at the rate of 0.2 A/s. The conduction layer provides charge dissipation during the EBL exposure. The sample was exposed to create a desired pattern using a 100 kV EBPG5000+ electron beam lithography system. The exposed resist was then developed in a mixture of methyl isobutyl ketone (MIBK) and IPA in the ratio of 1:3. The development was performed for 1 min and the sample was then rinsed in a fresh IPA and deionised water.

*Image analysis*

Image analysis was performed using custom LabVIEW code. The $T_1$ relaxation images were obtained by determining the $T_1$ decay curve at each pixel and fitting the data to a stretched exponential of the form $y = A \exp\left(\left(t/T_1\right)^p\right) + c$, where A is the amplitude of the exponential decay, $T_1$ is the spin lattice relaxation time, p is the stretched exponential power (p = 1 represents a single exponential decay) and c is the offset. Near surface NV defects are known to exhibit a distribution of $T_1$ times from the NV ensemble depending on their proximity to the surface and local spin environment [18]. This distribution leads to a non-exponential $T_1$ decay which is characterised well by a stretched exponential function. The amplitude and offset are left as free parameters since these can vary depending on the alignment of the magnetic field particularly around the excited state level anti-crossing (ESLAC) at 512 G.

*Magnetic field alignment*

Magnetic field alignment was achieved by monitoring the NV fluorescence signal near the ESLAC as described in [35]. The fluorescence at 512 G was extremely sensitive to the field alignment and could be aligned with a particular axis to a precision better than 0.1°. The magnetic field gradient is characterised across the field of view by determining the Zeeman splitting at each pixel via optically detected magnetic resonance (see SI). The measured field gradient in the y direction was 0.4% and 0.07% in the x direction.

*Continuous wave electron spin resonance*

Continuous-wave ESR spectra were acquired using a CMS8400 X-band (9.4 GHz) spectrometer (Adani, Belarus) fitted with a TE102 cavity and operating at a fixed receiver time constant of 100 ms and 100 kHz magnetic field modulation. Solution measurements were made at room temperature using a quartz flat cell (Wilmad, WG-808-Q).

## Acknowledgements

The authors acknowledge Dr. Nikolai Dontschuk and Dr. Stuart Earl for assistance with the E-beam lithography. This research was supported in part by the Australian Research Council Centre of Excellence for Quantum Computation and Communication Technology (Project number CE110001027). This work was also supported by the University of Melbourne through the Centre for Neural Engineering and the Centre for Neuroscience. L.C.L.H acknowledges support of the Australian Research Council under the Laureate Fellowship scheme (FL130100119). S.P. acknowledges the support from the NHMRC Fellowship scheme (1005050). P.S.D acknowledges the support from the ARC Future Fellowship scheme (FT130100204). S.C.D acknowledges the support from the ARC Future Fellowship scheme (FT11010019). D.A.S. acknowledges support from the Melbourne Neuroscience Institute Fellowship Scheme.

## Author contributions statements

D.A.S., L.T.H. and L.C.L.H. conceived the idea of the experiment. D.A.S. designed and constructed the quantum magnetic resonance microscope. R.R. and D.A.S. acquired the data with assistance from P.S.D., S.P. and P.M. in transition metal bio-chemistry. L.T.H., L.C.L.H. and D.A.S. developed the theoretical framework with input from P.S.D., S.P., S.C.D. and P.M. to describe the results. E.P. fabricated the image mask. S.C.D. performed CW ESR measurements. All authors discussed the results and participated in writing the manuscript.

## Additional information

The authors declare no competing financial interest.